# Magneto-memristive switching in a two-dimensional layer antiferromagnet

Hyun Ho Kim[1†,] Shengwei Jiang[2†], Bowen Yang[1], Shazhou Zhong[1], Shangjie Tian[3], Chenghe Li[3], Hechang Lei[3], Jie Shan[2], Kin Fai Mak[2], and Adam W. Tsen[1]

[1] H. H. Kim, B. Yang, S. Zhong, and A. W. Tsen
Institute for Quantum Computing, Department of Chemistry, and Department of Physics and Astronomy, University of Waterloo, Waterloo, Ontario N2L 3G1, Canada

[2] S. Jiang, J. Shan, and K. F. Mak
School of Applied and Engineering Physics, Department of Physics, and Kavli Institute for Nanoscale Science, Cornell University, Ithaca, New York 14853, USA

[3] S. Tian, C. Li, and H. Lei
Department of Physics and Beijing Key Laboratory of Opto-electronic Functional Materials & Micro-Nano Devices, Renmin University of China, Beijing 100872, China

[†] These authors contributed equally to this work.

# Abstract

Memristive devices whose resistance can be hysteretically switched by electric field or current are intensely pursued both for fundamental interest as well as potential applications in neuromorphic computing and phase-change memory. When the underlying material exhibits additional charge or spin order, the resistive states can be directly coupled, further allowing for electrical control of the collective phases. Here, we report the observation of abrupt, memristive switching of tunneling current in nanoscale junctions of ultrathin $CrI_3$, a natural layer antiferromagnet. The coupling to spin order enables both tuning of the resistance hysteresis by magnetic field, and electric-field switching of magnetization even in multilayer samples.

Two-dimensional (2D) magnetic semiconductors[1-3], such as $CrI_3$, have recently been shown to host a wealth of exotic properties including giant tunnel magnetoresistance[4-7], gate- and pressure-tuned interlayer magnetism[8-12], as well as terahertz spin wave excitations of possible topological nature[13-17]. In this report, we demonstrate that high electric field induces robust, hysteretic switching of the tunneling current in sub-micron graphene (Gr)/$CrI_3$/Gr junctions, characteristic behavior of a locally active memristor[18, 19]. Caused by the positive feedback of self-heating, the switching is furthermore abrupt, with speed limited only by that of the external circuit. Using nanoscale magneto-optical imaging, we additionally show that the resistive states are connected with the magnetic order in ultrathin $CrI_3$. This coupling gives rise to reciprocal magneto-electric effects—an external magnetic field can be used to widely tune the hysteresis of the resistive transition, while the electric field can also be used to control for net magnetization across the layers. Our results not only advance the functionality of 2D $CrI_3$, but may potentially be generalized to synthetic layer antiferromagnets for higher temperature operation.

Our devices consist of ultrathin $CrI_3$ tunnel barriers with few-layer Gr electrodes in a crossbar geometry and full encapsulation within hexagonal boron nitride (hBN). A simplified schematic and measurement circuit are shown in Fig. 1a, while discussion of the fabrication process in inert atmosphere can be found in Methods as well as our previous reports [6, 7, 15]. We have kept the junction area small (0.2-0.5$\mu$m$^2$) as we find such a geometry yields more robust device characteristics. Fig. 1b shows a key representative finding of this work taken from a 16-layer (16L) $CrI_3$ device at 1.4K. An optical image of the central region is shown in Supplementary Information (SI), Fig. S1. When slowly ramping voltage (0.03mV/s) up to a relatively high, critical value, $V_C^H$, the tunnel current abruptly increases by over a factor of two. Further increase in voltage traces out a new state with higher conductance. When ramping voltage down from this state, current abruptly decreases at a lower critical value, $V_C^L$, but returns to the original, lower conductance state. These critical voltages are insensitive to the ramping speed over five orders of magnitude (see SI, Fig. S2), and thus can be taken as intrinsic properties of the device. Their difference with respect to sweep direction gives rise to hysteresis, making this $CrI_3$ device a locally active tunnel memristor, sharing similar current-voltage characteristics with devices incorporating complex oxides[18, 19]. This switching behavior is furthermore robust and reproducible over a wide range of $CrI_3$ thicknesses (see SI, Fig. S3), although it diminishes with fewer $CrI_3$ layers, for reasons which we will later discuss.

In order to understand the nature of this switching, we have measured its change with both perpendicular (out-of-plane) and parallel (in-plane) magnetic fields as well as temperature, as shown in Figs. 2A, B, and C, respectively. Overall, we observe that with initially increasing field or temperature, the current transitions get pushed to lower voltages. In particular, when the field is applied perpendicular to the layers, $V_C^L$ and $V_C^H$ decrease at different rates, such that the hysteresis window substantially widens with increasing field. This effect is further consistent with voltage-dependent current vs. magnetic field measurements shown in SI, Fig. S4. Our finding here demonstrates that the switching behavior in $CrI_3$ can be broadly tuned with even moderate fields and differentiates it from other memristor materials that do not host magnetic order.

When temperature or field is increased further, no current jumps are observed down to zero voltage. Specifically, the jumps disappear when either 1) the temperature exceeds the critical temperature for magnetic ordering ($T_C$ ~ 45K at $B$ = 0), 2) the perpendicular field exceeds the critical field to fully polarize the antiparallel spins across the $CrI_3$ layers out-of-plane ($B_C^{\perp}$ ~ 2T at 1.4K), or 3) the parallel field exceeds the critical field to polarize the spins in-plane ($B_C^{\parallel}$ ~ 6T at 1.4K). These critical values are consistent with previous reports[5-7, 15]. Taken together, these results indicate that memristive switching only occurs in the magnetic ground state with antiferromagnetic (AFM) interlayer coupling, while neither the higher temperature, paramagnetic (PM) state nor the field-induced spin-parallel state exhibits switching behavior.

When the current jumps are observed, what then is the nature of the state with higher tunneling conductance? First, its temperature dependence shows insulating behavior similar to the ground state (see SI, Figs. S5), which rules out a metal-to-insulator transition observed in other memristive oxides[18, 19]. An electrically induced structural transition is another possibility, especially given that bulk $CrI_3$ exhibits rhombohedral layer stacking at these temperatures[20], while ultrathin samples show a monoclinic structure[11, 12, 21]. The former yields ferromagnetic (FM) coupling between the layers, while the latter gives rise to AFM coupling. In the absence of a structural transition in 2D $CrI_3$, both the field-induced, spin-parallel (FM-like) state at $B > B_c$ and the PM state near or above $T_c$ exhibit higher conductance than the AFM ground state due to spin-dependent tunneling effects[4-6, 14]. A transition to the former may potentially be caused by spin-transfer torque[22], while abrupt heating can cause transition to the latter.

In order to further our understanding, we performed magnetic circular dichroism (MCD) measurements on another 14L $CrI_3$ device using both laser and broadfield illumination[23]. These measurements sense the out-of-plane component of sample magnetization, and so can differentiate between the various scenarios above. While the former scheme produces higher sensitivity, the latter enables fast imaging with diffraction-limited spatial resolution below 400nm, which can be used to directly resolve the local magnetization at the junction. The experimental setups are described in the Experimental Section. An optical image of the device is shown in the lowest panel of Fig. 3a.

We first discuss MCD results with laser fixed on the center of the junction. In Fig. 3b, we show both current vs. voltage and MCD vs. voltage taken simultaneously at 3.5K for three different magnetic field levels: 0, 1.8, and 2.2T. The measurements at additional fields are shown in SI, Fig. S6. The MCD traces are plotted on an absolute scale, while the current traces have been offset for easy comparison with MCD. We observe that in the AFM state at zero field, a current jump produces almost no change in the MCD signal, which remains near zero for all voltages applied. For comparison, the MCD signal is relatively high in the FM-like state at 2.2T, corresponding to full spin polarization of the $CrI_3$ layers. The current also does not switch abruptly in this state as previously mentioned. These results indicate that the memristive switching seen in the ground state does not correspond to a transition towards spin-parallel interlayer coupling and rules out either a change in stacking or spin-transfer torque effects.

Nevertheless, we notice that for an intermediate field of 1.8T, transition into the higher conductance state coincides with an abrupt change in magnetization of nearly a factor of two, while further increase in voltage in this state leads to a continuous increase of the MCD signal. The hysteresis with voltage sweep direction also matches between MCD and current. These observations indicate that the overall magnetization of the $CrI_3$ layers can be directly switched and further continuously tuned by electrical means. We emphasize that this effect is different than control of magnetization in 2D $CrI_3$ by electric field or gating as previously reported[8-10]. First, in those devices, an electric field up to 1V/nm is generated in the absence of significant current flow, while current is required to induce the transition here as we will show. Second, gate-induced changes have only been reported for monolayer and bilayer $CrI_3$ and will decrease in efficiency with increasing sample thickness for even the same electric field as only the surface layers are impacted. In contrast, the current jumps seen here become more pronounced with increasing $CrI_3$ thickness (see SI, Fig. S3a), and so memristive switching can be used to control for the magnetization even in multilayers. At zero field and 1.4K, all the observed jumps furthermore occur near a much smaller critical electric field of 0.3V/nm (see SI, Fig. S3b), independent of sample thickness, which, in principle, implies that even samples of order 100nm thickness can be controlled by voltage levels readily achieved in the lab.

The high spatial resolution of the broadfield technique can be further used to image the local MCD electrically induced in the nanoscale junction. The upper panels of Fig, 3a show the MCD images taken with increasing voltage bias at 1.8T. No signal is observed below $V_C^H$; however, clear MCD signal can be seen centered at the junction above the transition. The full width at half maximum value of the MCD spot is measured to be 1.2μm just above $V_C^H$ and becomes larger as the voltage increases (see SI, Fig. S7). Similar overall behavior is seen on another 20L device (See SI, Fig. S8).

The observation that the spatial extent of the induced MCD change exceeds the size of the junction (700nm x 700nm) is consistent with the heating scenario. Specifically, we now show that the higher conductance state seen in transport at zero field corresponds to a transition towards the PM state with higher temperature. The magnetization of the PM state is not expected to be significantly different than that of the AFM ground state as oppositely polarized layers quench the net magnetization for multilayer samples. However, the two states can be distinguished by their magnetic susceptibility, or effectively, $d(\mathrm{MCD})/dB^{\perp}$. As MCD detects the out-of-plane magnetization in $CrI_3$, its change with perpendicular field can be expected to show the characteristic temperature dependence of an antiferromagnet, or increasing susceptibility with increasing temperature up to the transition temperature. On the same 14L device, we first measured junction MCD vs. field at several different temperatures in the low conductance state with 1.5V and the results are plotted in the left inset of Fig. 3c. While there is hysteresis with respect to the sweep direction, the MCD slope at zero field for either sweep clearly increases with increasing temperature as expected. The main left panel of Fig. 3c explicitly shows the temperature dependence of the susceptibility extracted from the MCD slope for downward field sweep and will serve as a reference for the effective temperature of the junction.

We next measured junction MCD vs. field at constant cryostat temperature but under different bias voltages. Results taken at 3.5K are shown in the right inset of Fig. 3c. We observe that the slope increases significantly when the voltage is above $V_C^H$ = 2.94V, indicating heating in the junction. The bias dependent MCD slope is plotted explicitly in the main right panel of Fig. 3c for several different temperatures. We observe that for each set temperature, the susceptibility is nearly constant at low bias, but increases substantially above $V_C^H$. Comparison with the left panel indicates that the effective temperature increases by ~20K regardless of the initial temperature. The result is an abrupt transition towards the PM state with higher tunneling conductance.

The totality of these effects can be summarized by the phase diagram shown in Fig. 4 for 1.4K operation, with electric field serving as a nonlinear control parameter for junction temperature. The data points are extracted from the plots in Fig. 2a and b. Under zero magnetic field, increasing electric field across the junction beyond ~0.3V/nm drives $CrI_3$ from the layer AFM ground state towards the higher temperature PM state. Applying magnetic field in either the out-of-plane or in-plane direction, both the initial and final states upon driving can be tuned. In particular, a small perpendicular field below $B_C^{\perp}$ ~ 2T does not substantially alter the AFM ground state but stabilizes an FM-like final state with net out-of-plane magnetization. The transition furthermore becomes more hysteretic as denoted by the green region. On the other hand, a parallel field below $B_C^{\parallel}$ ~ 6T both creates canted AFM (cAFM) spins initially and stabilizes an FM-like final state with net in-plane magnetization. Applying even larger magnetic fields terminates the electrically driven transition and fully aligns the spins along the field direction. The memristive switching behavior in ultrathin $CrI_3$ is thus highly tunable and closely coupled to the various magnetic states.

Such a change is not easily anticipated as Joule heating in devices is typically expected to be a gradual process. Instead, the timescale of the heating transition in our $CrI_3$ junction is below 40ns as we will show in Fig. 5. Fast thermal switching has been previously observed in various nonlinear solid-state devices induced by the positive feedback of self-heating[24-26]. In this scenario, initial heating under a fixed voltage bias will cause an increase in device conductance, which then leads to enhanced current triggering more heating. This instability can abruptly induce a thermal transition and is likely to be the underlying cause for the memristive switching in $CrI_3$. In particular, near the critical electric field within the AFM state, the current increases sharply with increasing temperature, providing the positive feedback necessary to drive the transition. This effect is reduced in thinner samples due to the exponential dependence of spin-dependent tunneling current on sample thickness[5, 7]. In contrast, in the PM state above $T_C$ current increases with temperature more gradually, while the FM-like state above $B_C$ shows the opposite trend of decreasing current with increasing temperature. In both cases, the self-heating becomes quenched. A comparison of the temperature dependence for the three states can be found in SI, Figs. S5. Finally, at finite magnetic fields near $B_C$, heating will cause a transition to the FM-like (instead of PM) state with larger net magnetization as the critical fields decrease with increasing temperature[4, 6, 8].

In order to demonstrate the performance of our memristor, we inputted a series of voltage waveforms on our devices at 1.4K and 0T and examined the current output from the $CrI_3$ junction.

These results are shown in Fig. 5. For the first 16L device, we see that a single voltage level (~3.33V) applied within the hysteresis region can yield either low or high current states depending on the input history: whether the previous voltage level was <$V_C^L$ or >$V_C^H$, respectively (Fig. 5a). The current response is further robust over many cycles. On the timescale of several seconds shown here, the switching appears to be instantaneous. In order to measure the speed of the switching, we next introduced a fast voltage pulse with ~100ns rise and fall time and observed the current response on another 12L device (Fig. 5b). When the voltage level exceeds $V_C^H$, the heating transition associated with the current rise occurs within ~40ns, while the cooling (falling) transition is ~100ns. We note that the circuit has not been optimized for high frequency operation, and so these values should be taken as upper limits. Nevertheless, these results indicate that MHz response frequencies should be obtainable even with our existing setup, and so in Fig. 5c we monitored the current response of the 12L device to a 1MHz sinewave voltage with fixed amplitude (0.1V) and changing DC offset. When the offset is chosen so that no transition is made, the current output is also purely sinusoidal with low or high amplitude depending on whether the device is in the low or high conductance state, respectively. Across the transition voltage, faster dynamics can be resolved corresponding to the switching between the two states. Finally, the application of a magnetic field can lead to oscillations in both current and MCD signal (see SI, Fig. S9), allowing for the possibility of fast electrical control of magnetization in thin samples.

Our work demonstrates a nanoscale memristor device with simultaneous magnetic field controllable resistance switching and electric field controllable magnetization. Although operation is currently restricted to low temperature, the thermally induced mechanism is completely general and may be applicable to a wide range of materials. In particular, synthetic antiferromagnets[27] with layer-dependent magnetic ordering similar to $CrI_3$, but higher $T_c$, could potentially be substituted for operation at elevated temperatures. Our device concept further enhances the functionality of 2D magnets and provides a new approach for future explorations in neuromorphic computing and ultrafast memory.

## Experimental Section

**Crystal synthesis.** $CrI_3$ single crystals were grown by the chemical vapor transport method. $CrI_3$ polycrystals were placed inside a silica tube with 200mm length and 14mm inner diameter. After evacuation to 0.01Pa and sealing, the tubes were moved into a two-zone horizontal furnace. The temperature of the source (growth) zone was slowly raised to between 873-993K (723-823K) over a 24h period, and then held there for 150h.

**Device fabrication.** All fabrication process was carried out inside a nitrogen-filled glove box ($P_{O_2}$, $P_{H_2O}$ < 0.1ppm) to minimize sample degradation. Graphite (Gr) from CoorsTek, hexagonal boron nitride (h-BN) from HQ graphene, and $CrI_3$ were exfoliated on Si wafers with 285-nm-thick $SiO_2$. By utilizing a conventional pickup technique using poly(bisphenol A carbonate) (PC), h-BN/Gr/$CrI_3$/Gr/h-BN structures were sequentially stacked onto a polymer stamp and then transferred onto another $SiO_2$/Si chip having pre-patterned Au(40nm)/Ti(5nm) electrodes contacting the Gr. We chose thin (<5nm) and narrow (500-700nm) Gr flakes. The Gr/$CrI_3$/Gr junction area was kept between 0.25-0.5μm$^2$. One such device is shown in Fig. S1.

**Transport measurements.** Magnetotransport measurements were performed in a He4 cryostat (base temperature 1.4K). DC current/voltage measurements were performed with a Keithley 2450 source measure unit. A Keysight 33210A function generator and a Tektronix AWG5014B oscilloscope were used for high-frequency measurements, and the $CrI_3$ output current was measured by the voltage across a 50Ω series resistor.

**Magneto-optical measurements.** Magnetic circular dichroism (MCD) measurements were performed in an attoDry1000 cryostat (base temperature 3.5K). For point measurements, a HeNe laser at 632.8 nm was used. The laser beam was coupled into and out of the cryostat using free-space optics. The incident laser beam was focused onto the sample by a cryogenic objective. The beam size on the sample was ~1μm$^2$ and the power ~5μW. The incident beam was modulated between left and right circular polarization using a photoelastic modulator at 50.1kHz. The reflected beam was collected by the same objective and detected by a photodiode. The MCD signal was determined as the ratio between the AC component of the reflected light intensity at 50.1 kHz (measured with a lock-in amplifier) and the DC component (measured with a digital multimeter). For imaging, broadfield illumination from an incoherent white LED was sent through a bandpass filter centered at 632nm. A linear polarizer and a rotational quarter wave plate were used to generate left and right circularly polarized light. A liquid-nitrogen-cooled CCD detector was used to detect the reflected light. The MCD image was calculated as the difference between the left and right circularly polarized light reflection intensity normalized by the total reflection intensity. The spatial resolution is diffraction limited (~400 nm).

## Acknowledgements

We thank Youngki Yoon, Demin Yin, and Guo-Xing Miao for useful discussions, as well as Jonathan Baugh for lending us the high-frequency electronics. AWT acknowledges support from the US Army Research Office (W911NF-19-10267), an Ontario Early Researcher Award (ER17-13-199), and the Korea-Canada Cooperation Program through the National Research Foundation of Korea (NRF) funded by the Ministry of Science, ICT and Future Planning (NRF-2017K1A3A1A12073407). This research was undertaken thanks in part to funding from the Canada First Research Excellence Fund. The magneto-optical measurements at Cornell were supported by NSF (DMR-1807810) and ONR (award N00014-18-1-2368). HL acknowledges support from the National Key R&D Program of China (Grants No. 2016YFA0300504), the National Natural Science Foundation of China (No. 11574394, 11774423, 11822412), the Fundamental Research Funds for the Central Universities, and the Research Funds of Renmin University of China (RUC) (15XNLQ07, 18XNLG14, 19XNLG17).

## Competing interests

The authors declare no competing interests.

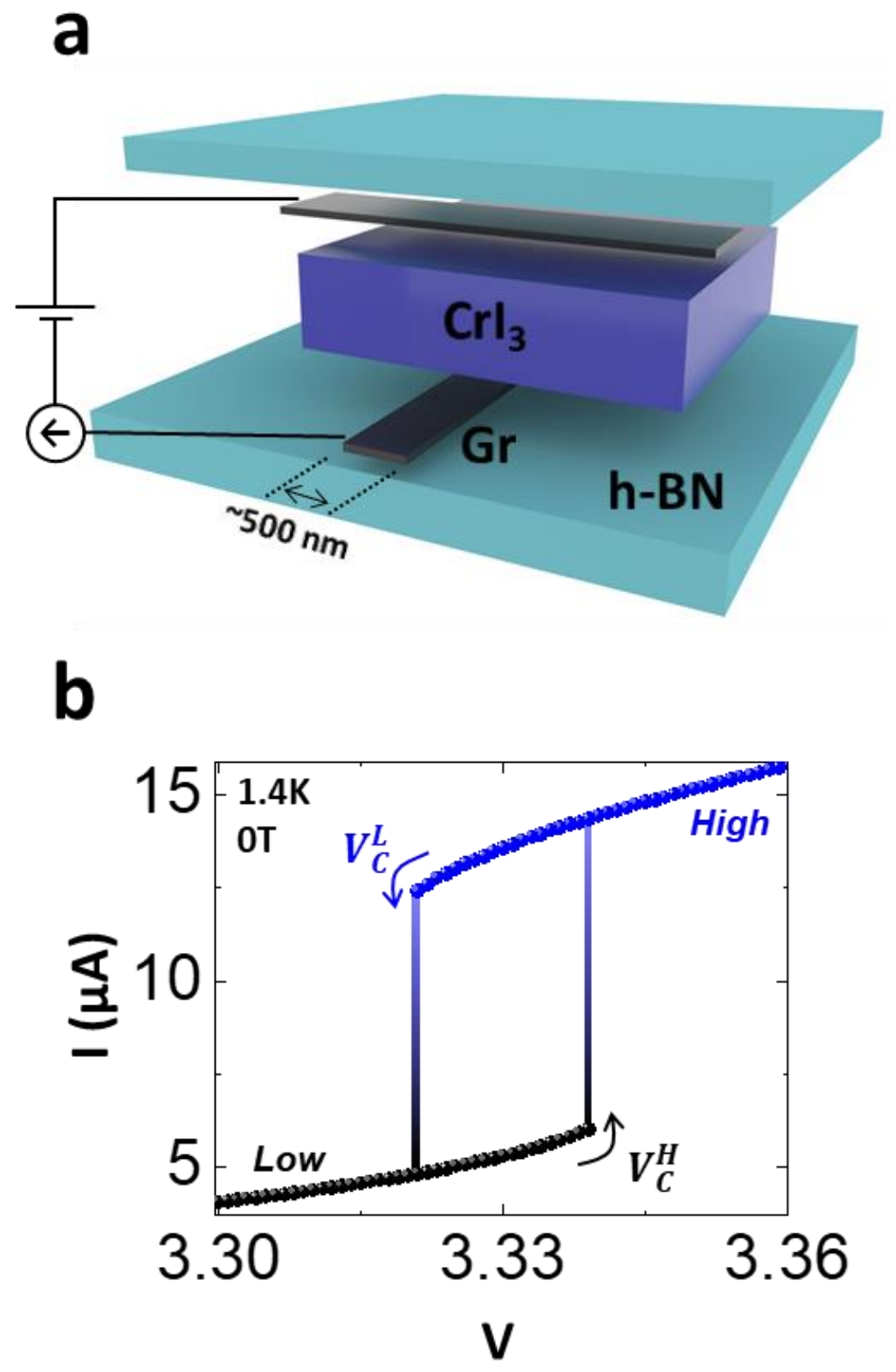

**Figure 1. Memristive switching in nanoscale $CrI_3$ tunnel junctions. (a)** Schematic illustration of the device. **(b)** Observation of abrupt and hysteretic switching of tunneling current across 16L $CrI_3$ at 0T at 1.4K.

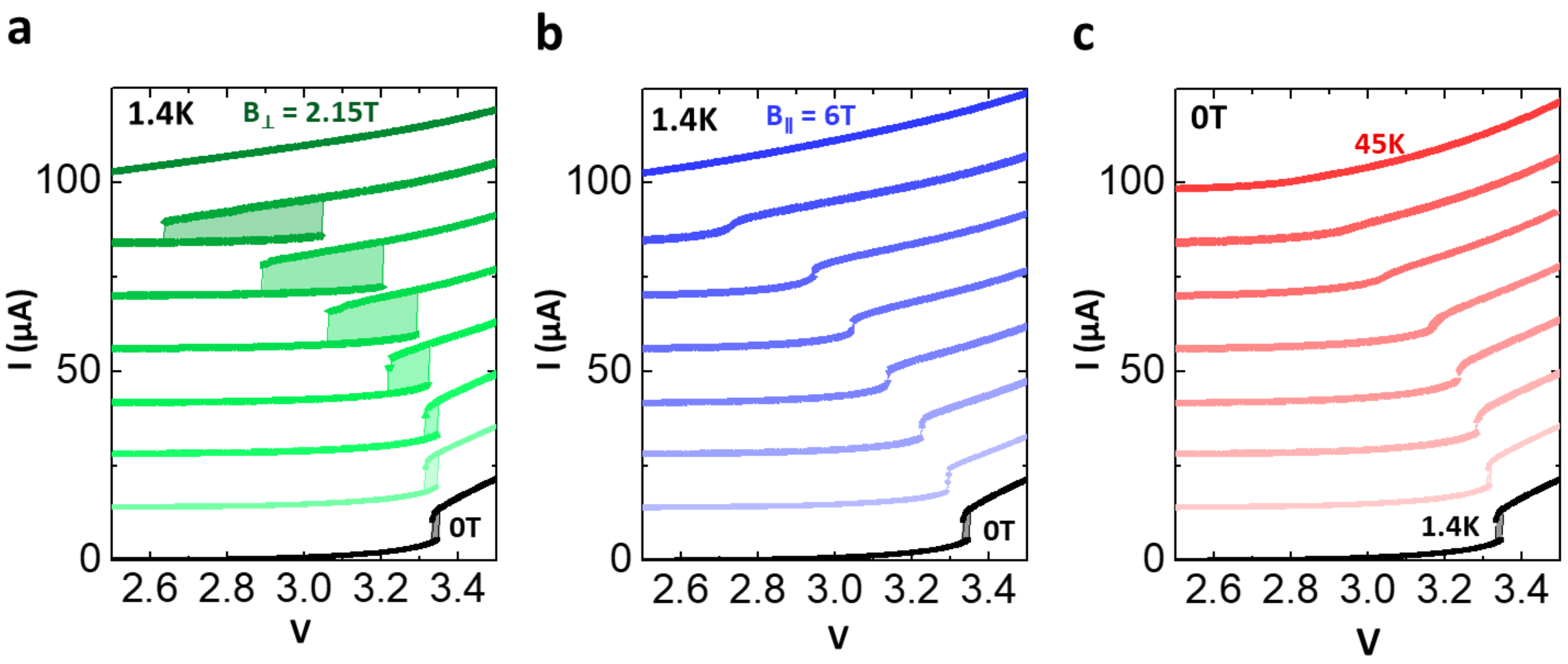

**Figure 2. Magnetic field and temperature dependence of memristive switching.** Current vs. voltage sweeps for several different **(a)** perpendicular magnetic field levels (0, 0.5, 1, 1.2, 1.5, 1.65, 1.8, and 2.15T) **(b)** parallel magnetic field levels (0, 1, 2, 2.5, 3, 4, 5, and 6T) and **(c)** temperatures (1.4, 15, 20, 25, 30, 35, 40, and 45K) from bottom to top, respectively. Current levels are offset by 14µA for easy comparison.

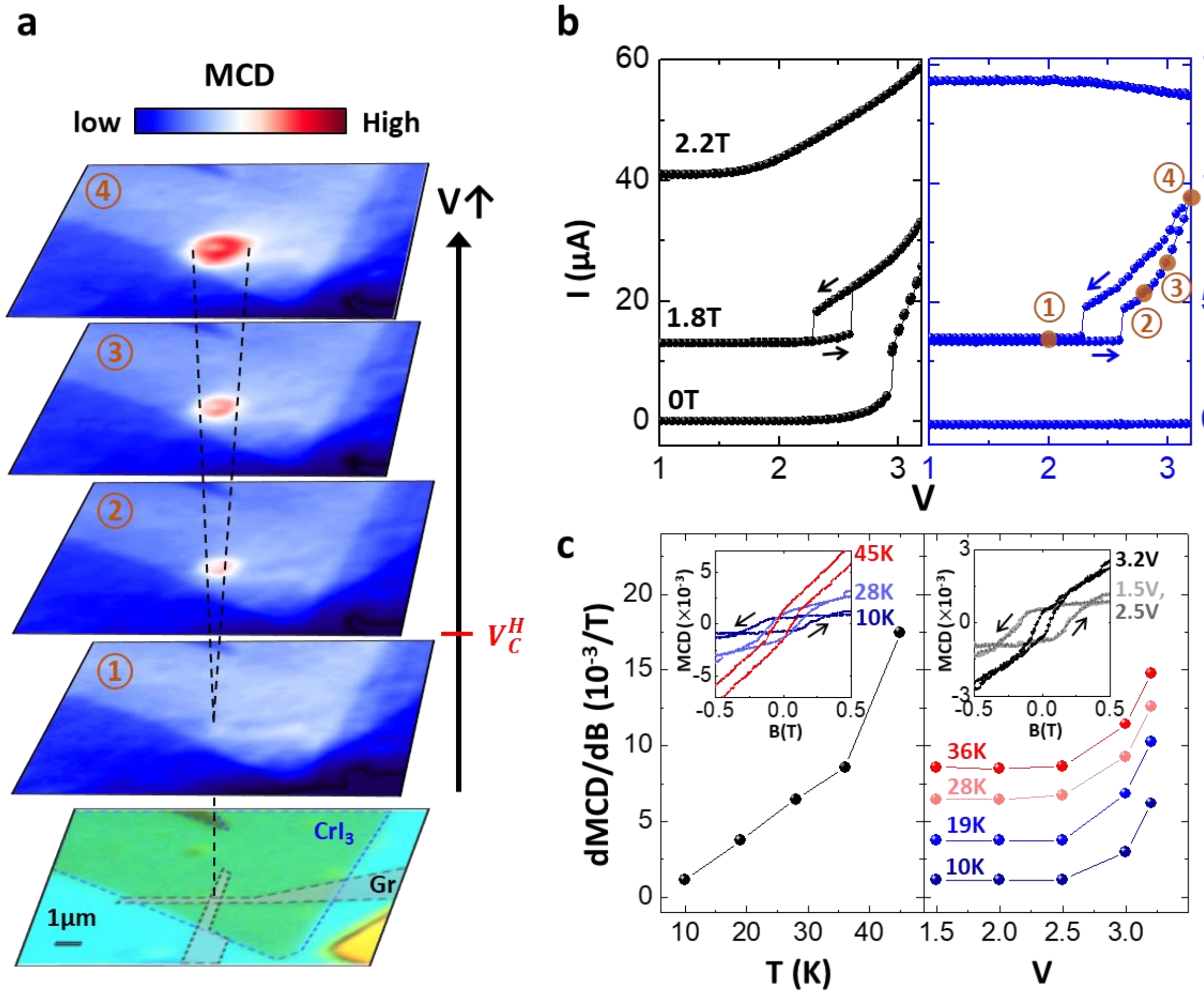

**Figure 3. Combined electrical and magneto-optical measurements across the switching transition. (a)** Voltage-dependent magnetic circular dichroism (MCD) microscopy on a 14L $CrI_3$ device taken by broadfield illumination at 3.5K and $B_{\perp}$= 1.8T. Minimum resolution is ~400nm. **(b)** Simultaneous current and MCD measurement as a function of voltage at different perpendicular magnetic field levels with laser fixed on junction. The numbers in the right panel show voltage conditions for **(a).** Current levels at 1.8T and 2.2T are offset for easy comparison. **(c)** Temperature- and voltage- dependent $d\mathrm{MCD}/dB_{\perp}$. Insets in **(c)** show MCD vs $B_{\perp}$ with different temperature (10, 28, and 45K) and voltage conditions (1.5, 2.5, and 3.2V), from which the slopes are extracted.

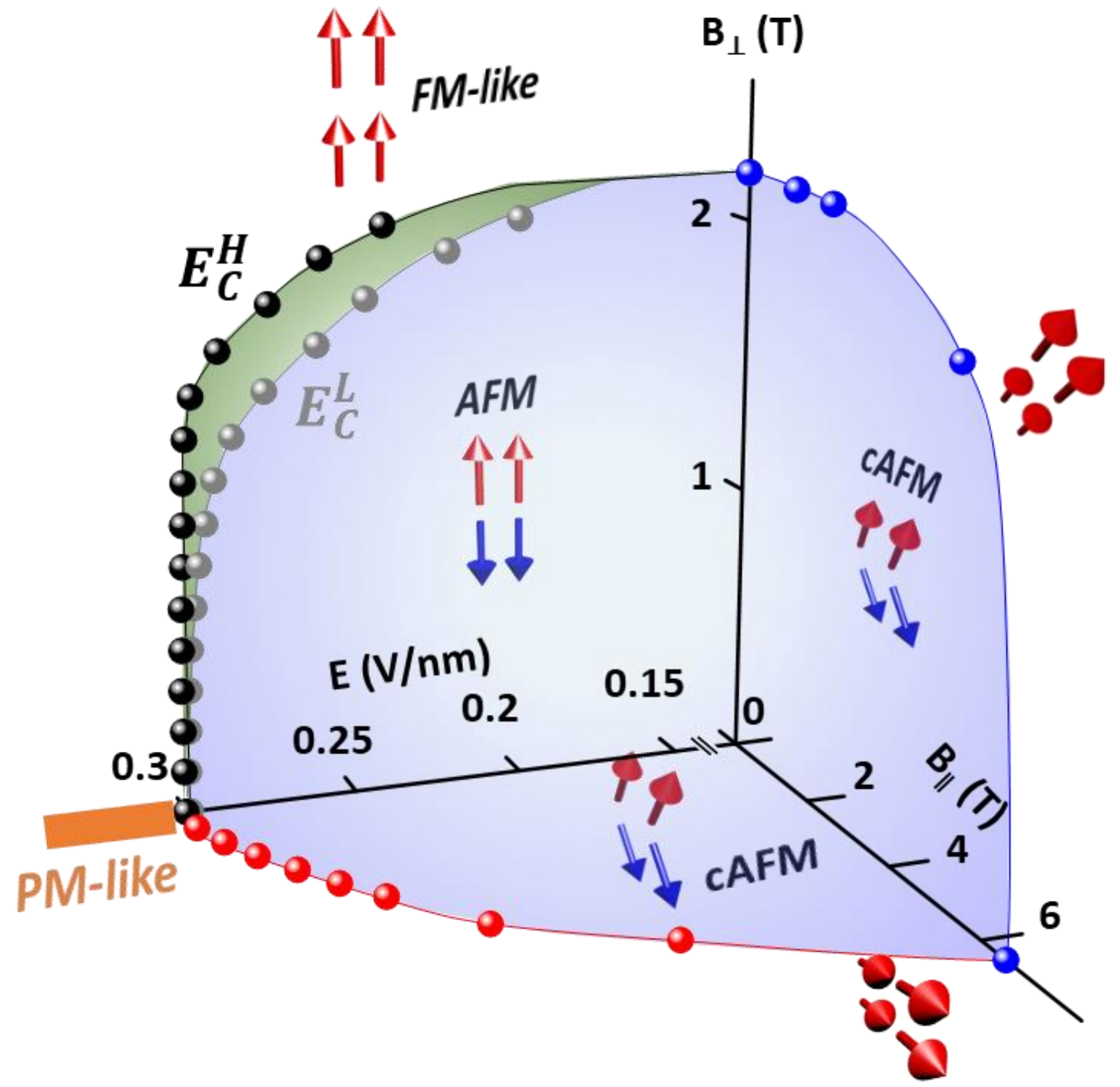

**Figure 4. Three-dimensional phase diagram of magnetic states in 2D $CrI_3$ at 1.4K under electric and magnetic fields.**

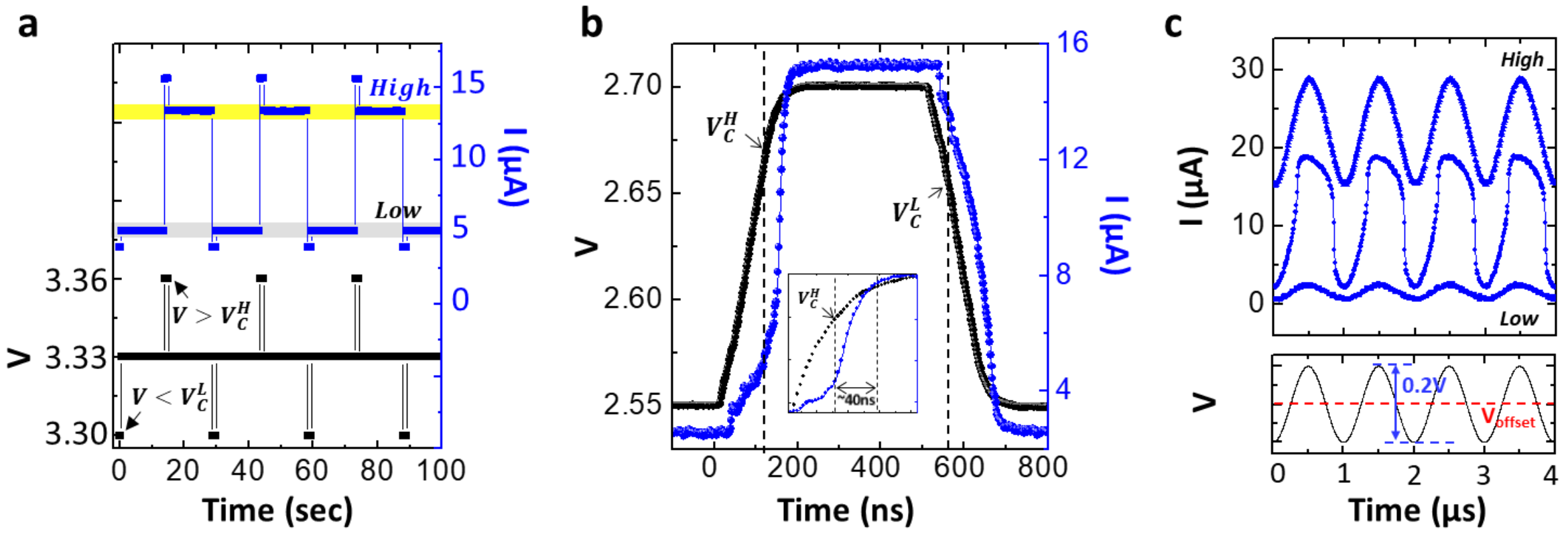

**Figure 5. Fast and robust switching behavior of $CrI_3$ memristor. (a)** Current measurement in 16L $CrI_3$ device with pulsed voltage waveforms at 1.4K and 0T, showing two current states at 3.33V depending on voltage history. **(b)** Current response with a fast voltage pulse in 12L $CrI_3$ device at 1.4K and 0T. Inset shows a zoom-in for the rising transition. Heating (cooling) transition is circuit-limited at ~40ns (~100ns). **(c)** Current response to a 1MHz sinewave voltage with fixed amplitude (0.1V) and different offset voltages (2.4V for low, 2.63V for across the transition, and 2.8V for high) in 12L device.

## Supporting Information

# Magneto-memristive switching in a two-dimensional layer antiferromagnet

Hyun Ho Kim[1†,] Shengwei Jiang[2†], Bowen Yang[1], Shazhou Zhong[1], Shangjie Tian[3], Chenghe Li[3], Hechang Lei[3], Jie Shan[2], Kin Fai Mak[2], and Adam W. Tsen[1]

[1]*Institute for Quantum Computing, Department of Chemistry, Department of Physics and Astronomy, and Department of Electrical and Computer Engineering, University of Waterloo, Waterloo, Ontario N2L 3G1, Canada*

[2]*School of Applied and Engineering Physics, Department of Physics, and Kavli Institute for Nanoscale Science, Cornell University, Ithaca, New York 14853, USA*

[3]*Department of Physics and Beijing Key Laboratory of Opto-electronic Functional Materials & Micro-Nano Devices, Renmin University of China, Beijing 100872, China*

[†] These authors contributed equally to this work.

## I. An example optical image of a completed device

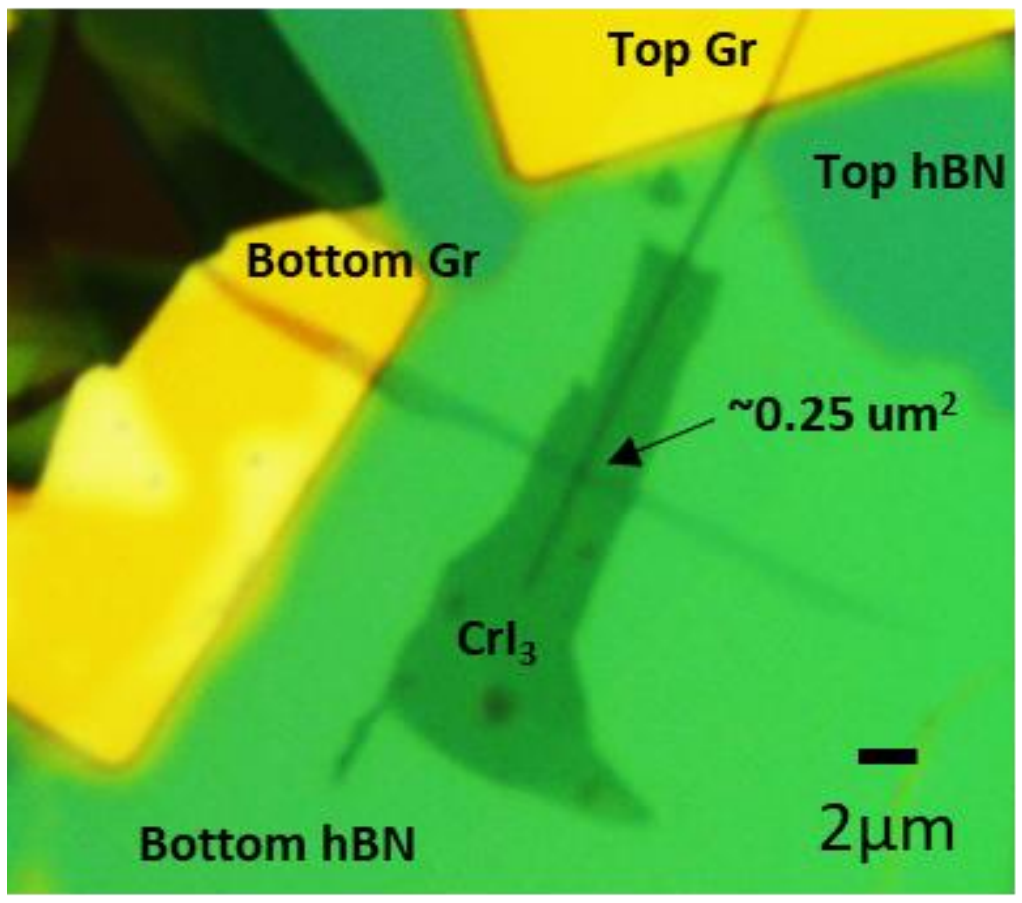

**Figure S1.** Optical image of h-BN/Gr/16L $CrI_3$/Gr/hBN structure used for data in Figs. 1b and 2. Each Gr is connected to a pre-patterned Au electrode. The $CrI_3$ flake is fully encapsulated by h-BN. Gr/$CrI_3$/Gr junction area is ~0.25μm$^2$.

## II. Sweep rate dependence of critical voltages

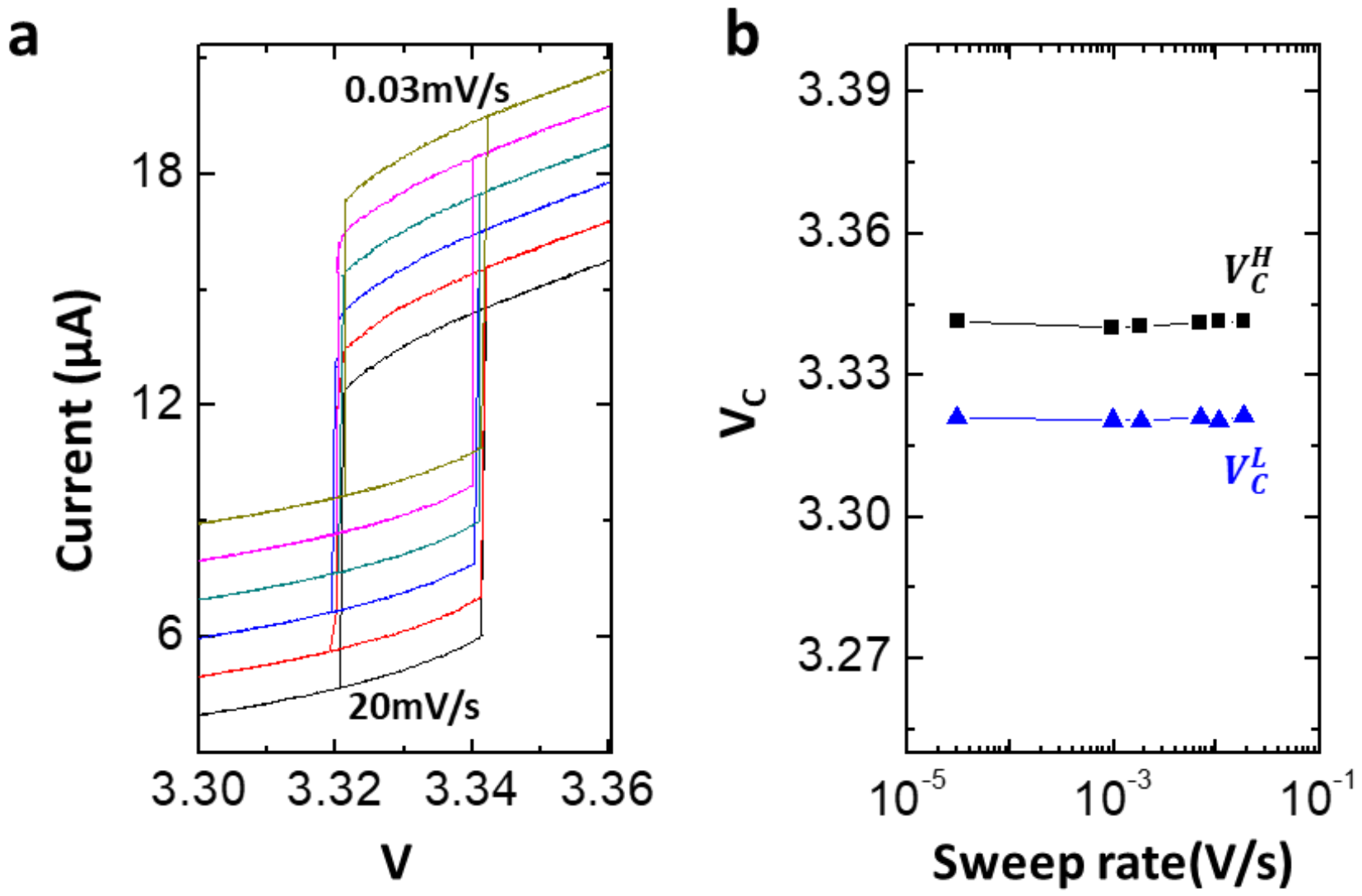

**Figure S2.** Sweep rate dependence in 16L $CrI_3$ (same device used for Figs. 1b and 2). **(a)** I-V plots showing switching behavior for 0.03, 1, 2, 7, 10, and 20 mV/s voltage sweep rates, in sequence from top to bottom. Current levels are offset by 1μA for easy comparison. **(b)** $V_C^H$ and $V_C^L$ vs. sweep rate.

## III. Thickness dependence of switching behavior

In Fig. S3a, we show the J-V plots for 3-20L $CrI_3$ devices. Overall, the magnitude of the current jump increases with the number of layers. The jump cannot be observed below 4 layers. Furthermore, all the jumps are observed to occur near 0.3V/nm critical electric field, independent of thickness, as shown in Fig. S3b.

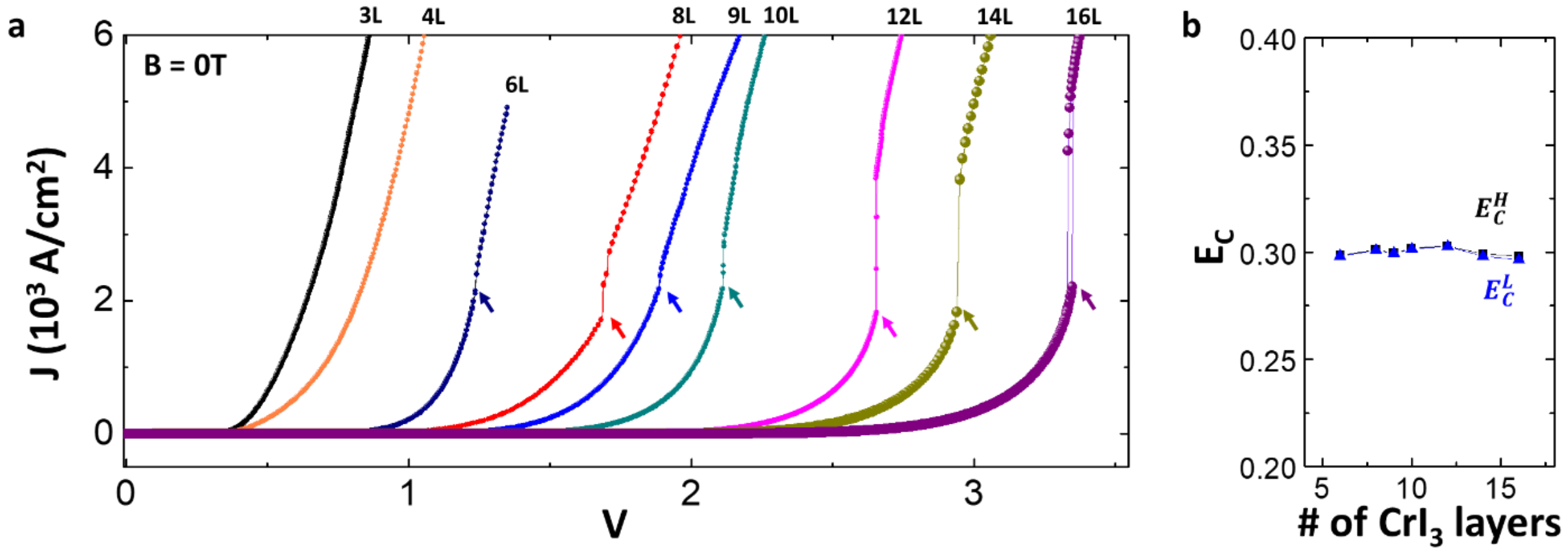

**Figure S3. (a)** Current density vs. voltage as a function of the number of $CrI_3$ layers at 1.4K and 0T. **(b)** Critical electric field vs layer number.

## IV. Additional magnetotransport data

We additionally measured current vs. magnetic field at different voltage levels with the same 16L device used for Fig 1b and 2, as shown in Fig. S4a (out-of-plane field) and S4b (in-plane field). For perpendicular field, the critical field for reaching the higher conductive state decreases from 2T as the voltage increases. The abrupt jumps disappear when $V \geq V_C^H = 3.34\text{V}$. For in-plane field, critical field also decreases with increasing voltage; however, the jumps are less abrupt at lower voltages.

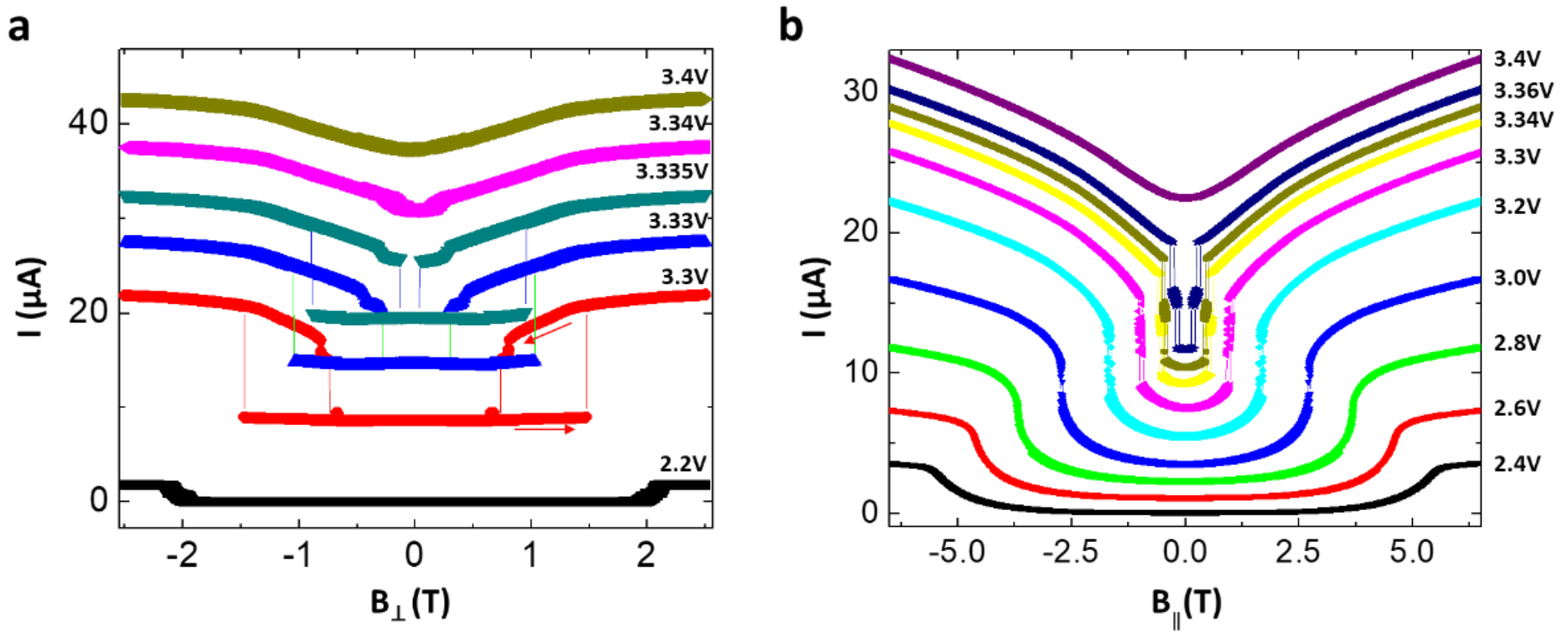

**Figure S4.** Voltage-dependent current vs. B field in 16L $CrI_3$ device for **(a)** out-of-plane and **(b)** in-plane field orientation.

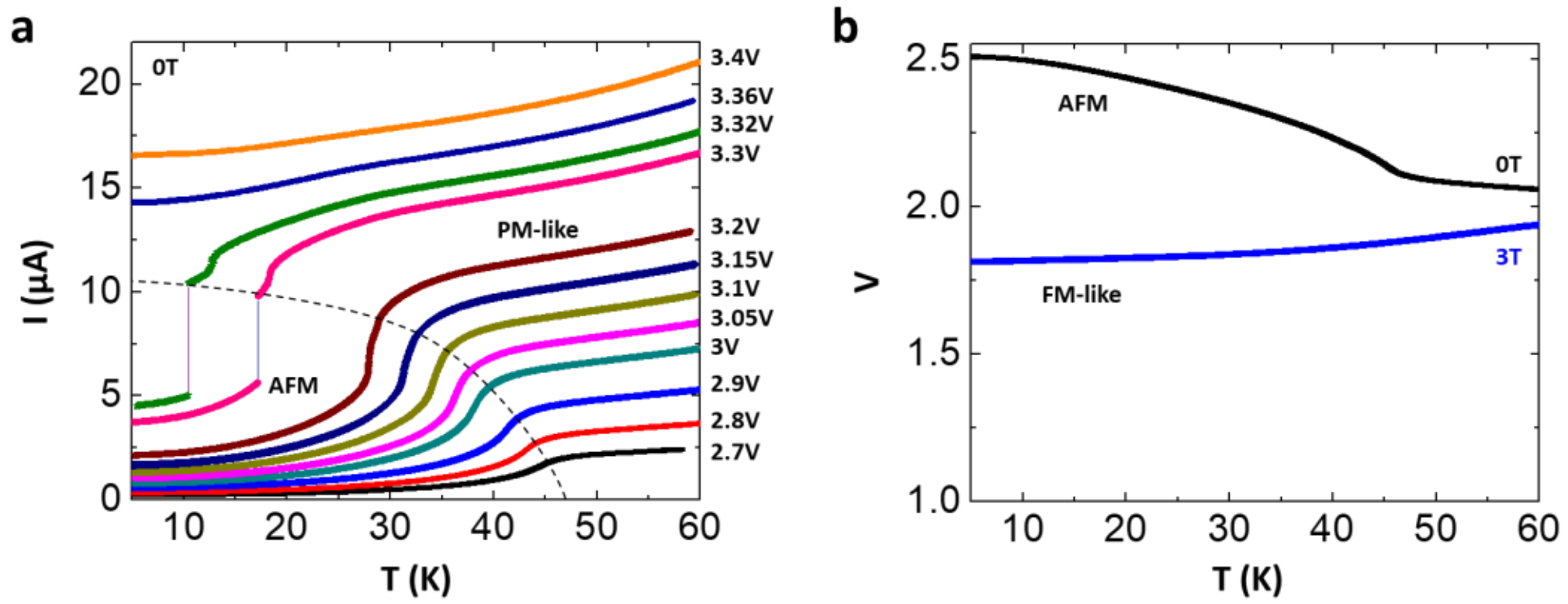

**Figure S5. (a)** Zero-field current vs. temperature for 16L $CrI_3$ device at different voltage levels all showing insulating behavior. The voltage levels where current jumps are observed correspond to $V_C^L$ at the jump temperature. The same device is used for Figs. 1b and 2. Temperature cooling rate is ~1K/min. **(b)** Voltage vs. temperature under 2μA current biasing for 12L $CrI_3$ device for two magnetic field levels.

## V. Additional MCD data

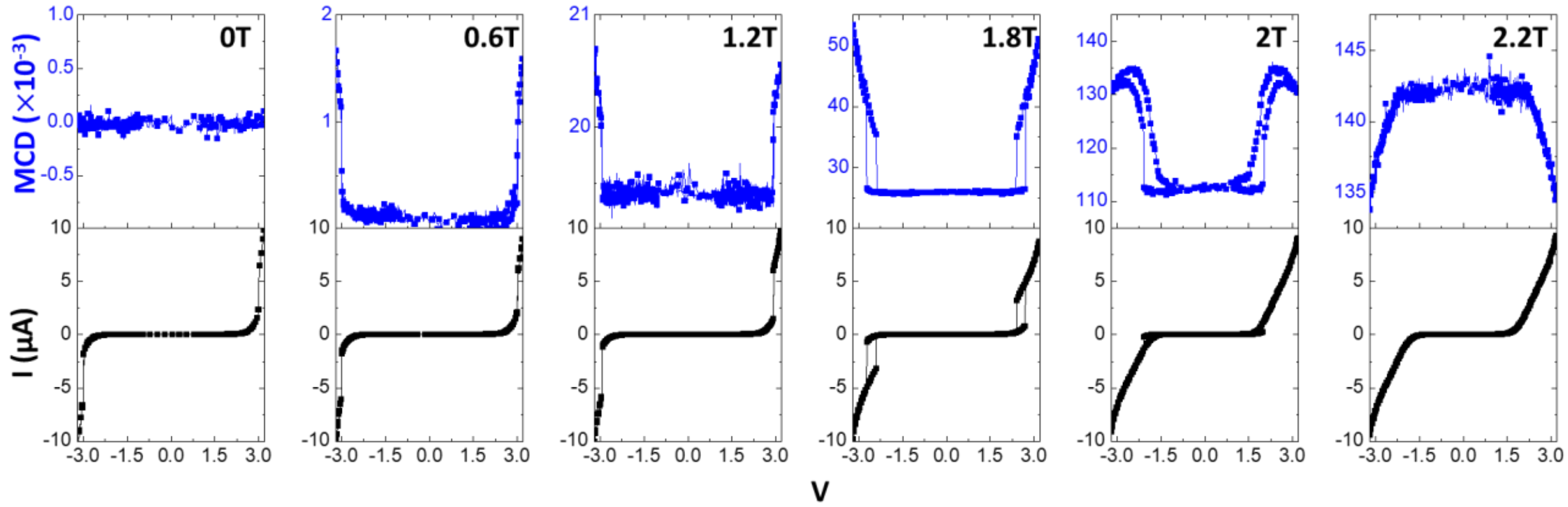

**Figure S6.** Simultaneous MCD and current measurements as a function of voltage for different perpendicular magnetic field levels at 3.5K. The same device is used for Fig. 3b.

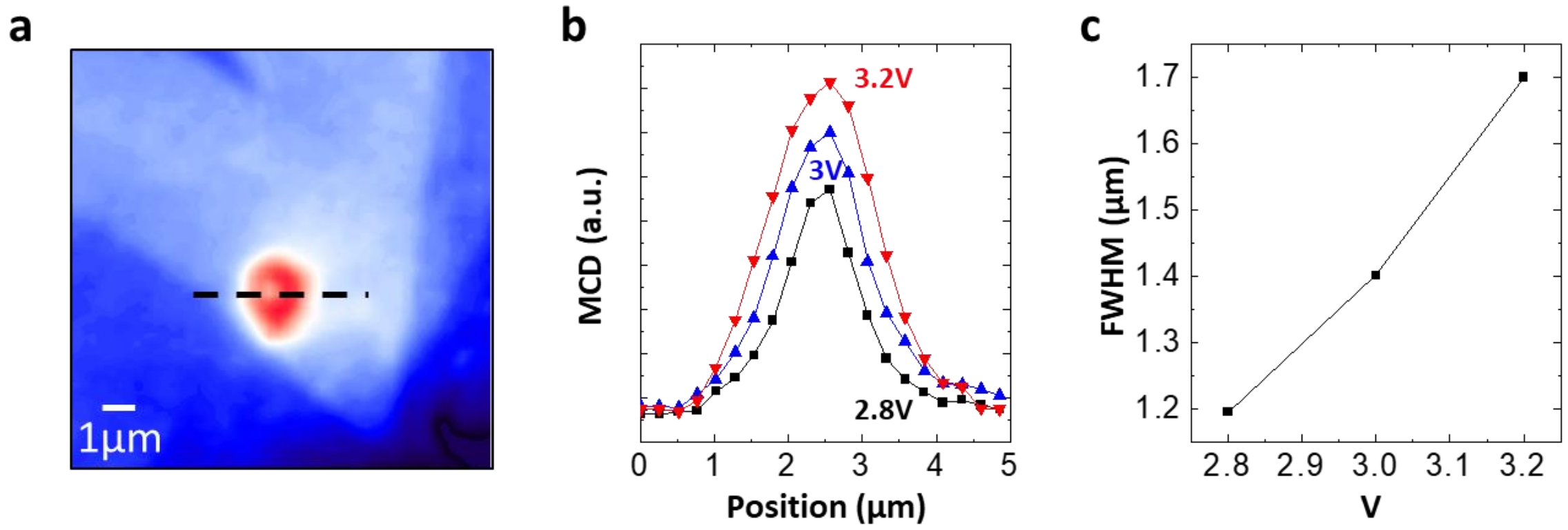

**Figure S7.** Determination of the MCD spot size. **(a)** Example MCD image taken at 3.2V, 1.8T, and 1.4K. The same image is shown in the top panel of Fig. 3a. **(b)** Cross-sectional profiles of the MCD signal measured along the dashed line in A as a function of voltage. **(c)** Full width at half maximum (FWHM) of the MCD peaks in B vs. voltage.

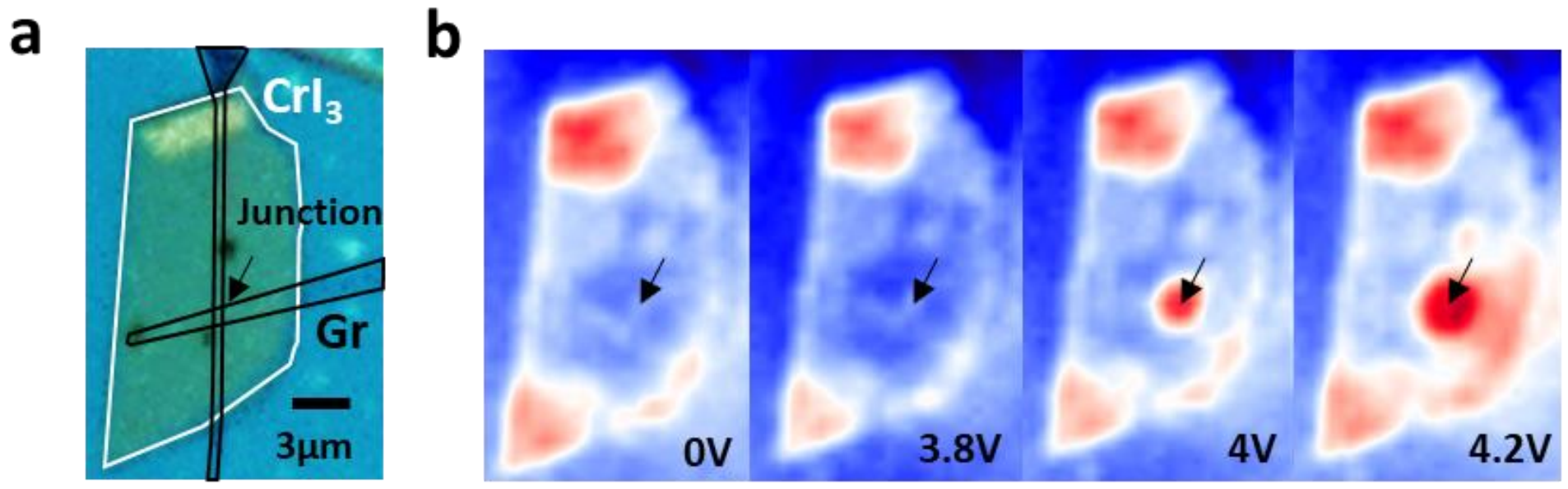

**Figure S8.** Additional MCD imaging of ~20L $CrI_3$ device showing measurement reproducibility. (a) An optical image of the device and (b) voltage-dependent magnetic circular dichroism (MCD) microscopy.

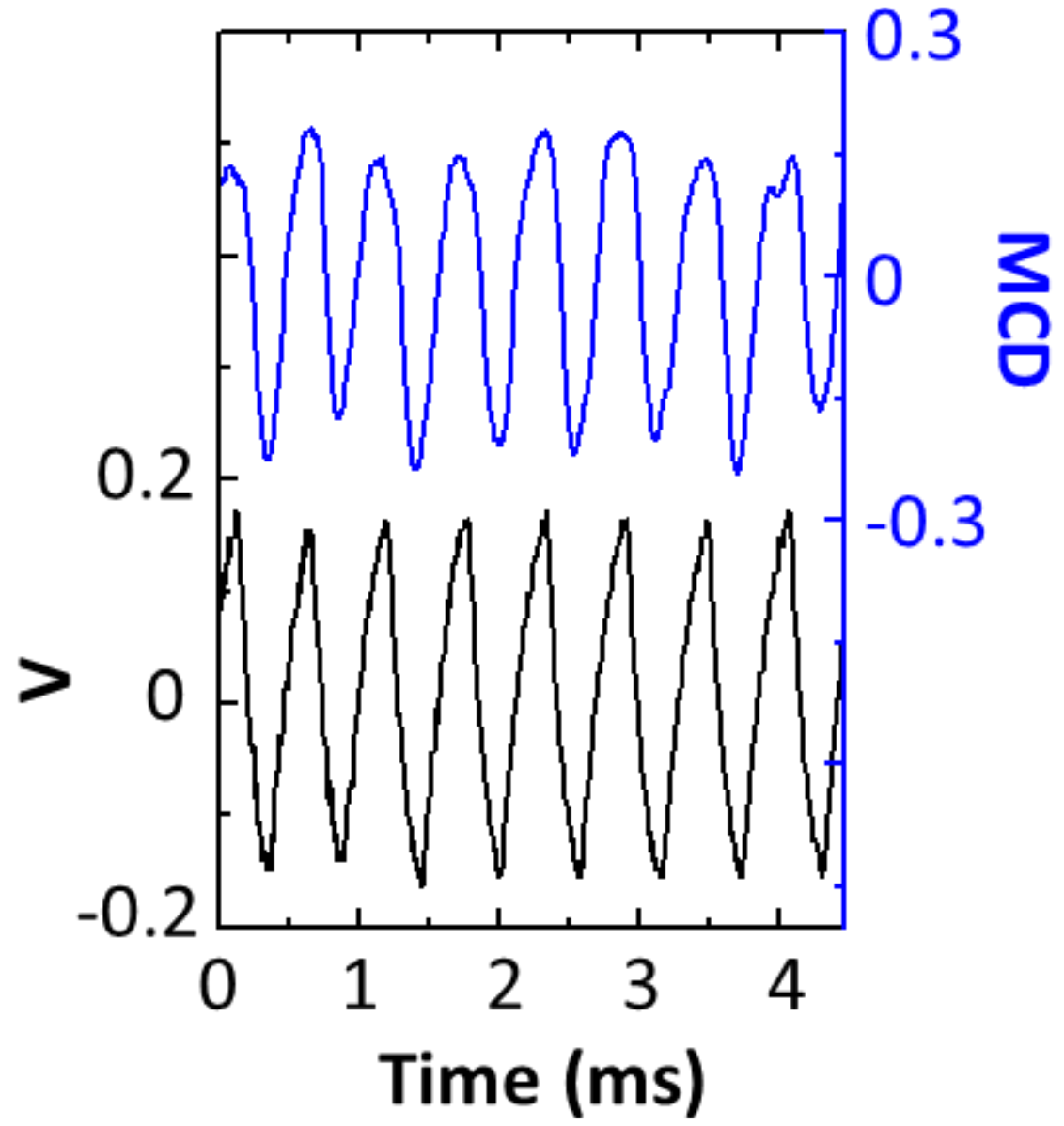

**Figure S9.** Simultaneous oscillations of voltage and MCD at 1.8T. A DC current of 3μA is applied between the high and low current states in order to establish voltage oscillations.